\documentclass[10pt]{iopart}
\usepackage{graphicx}
\usepackage{tabls}
\eqnobysec

\newcommand{\beq}{\begin{equation}}
\newcommand{\beqa}{\begin{eqnarray}}
\newcommand{\eeq}{\end{equation}}
\newcommand{\eeqa}{\end{eqnarray}}

\newcommand{\ket}[1]{\vert#1\rangle}
\newcommand{\braket}[2]{\langle#1\vert#2\rangle}
\newcommand{\braopket}[3]{\langle#1\vert#2\vert#3\rangle}

\newcommand{\abs}[1]{\left\vert#1\right\vert}
\newcommand{\dd}{{\rm d}}
\newcommand{\dif}[2]{{\frac{\dd#1}{\dd#2}}}
\newcommand{\eps}{\varepsilon}
\newcommand{\etc}{\!\!\!\cdots}
\newcommand{\frad}[2]{{\displaystyle{\displaystyle#1\over\displaystyle#2}}}
\newcommand{\fras}[2]{{\scriptstyle{\scriptstyle#1\over\scriptstyle#2}}}
\newcommand{\h}{\widehat}
\newcommand{\half}{{\fras{1}{2}}}
\newcommand{\ii}{{\rm i}}
\newcommand{\p}{\psi}
\newcommand{\q}{{\bf q}}
\newcommand{\sgn}{\mathop{\rm sgn}}
\newcommand{\vv}{{\vphantom{X}}}
\newcommand{\w}{^{({\rm w})}}
\newcommand{\ct}{^{({\rm cont})}}
\newcommand{\wct}{^{({\rm w,cont})}}
\renewcommand{\H}{{\cal H}}
\newcommand{\I}{{\cal I}}
\newcommand{\RR}{{\overline R}}

\begin{document}

\title[Quantum return probability of $N$ non-interacting lattice fermions]
{Quantum return probability of a system\\ of $N$ non-interacting lattice fermions}

\author{P L Krapivsky$^{1,2}$, J M Luck$^2$ and K Mallick$^2$}

\address{$^1$ Department of Physics, Boston University, Boston, MA 02215, USA}

\address{$^2$ Institut de Physique Th\'eorique, Universit\'e Paris-Saclay, CEA and CNRS,
91191 Gif-sur-Yvette, France}

\begin{abstract}
We consider $N$ non-interacting fermions performing continuous-time
quantum walks on a one-dimensional lattice.
The system is launched from a most compact configuration
where the fermions occupy neighboring sites.
We calculate exactly the quantum return probability
(sometimes referred to as the Loschmidt echo) of observing
the very same compact state at a later time $t$.
Remarkably, this probability depends
on the parity of the fermion number -- it decays as a power of time for even $N$, while
for odd $N$ it exhibits periodic oscillations modulated by a decaying power law.
The exponent also slightly depends on the parity of $N$,
and is roughly twice smaller than what it would be in the continuum limit.
We also consider the same problem, and obtain similar results,
in the presence of an impenetrable wall at the origin
constraining the particles to remain on the positive half-line.
We derive closed-form expressions for the amplitudes
of the power-law decay of the return probability in all cases.
The key point in the derivation is the use of Mehta integrals,
which are limiting cases of the Selberg integral.
\end{abstract}

\eads{\mailto{pkrapivsky@gmail.com},\mailto{jean-marc.luck@ipht.fr},\mailto{kirone.mallick@ipht.fr}}

\maketitle

\section{Introduction}

The advent of techniques to manipulate cold atoms
has led to the experimental realization of low-dimensional
quantum gases~\cite{Dalibard} that,
for many decades, were mainly thought of as being toy models
for theoreticians~\cite{Paredes,Kinoshita}.
The continuous tuning of the pair interaction between atoms thanks to
Feshbach resonances allows one to create many-body systems
ranging from free to strongly interacting.
This revolution triggered a profusion of studies of the non-equilibrium dynamics
of quantum systems~\cite{Cazalilla,Polkovnikov,Eisert},
including quenches, entanglement and thermalization,
that resulted in many new ways of exploring the quantum world~\cite{Haroche}.

Non-interacting tight-binding fermions on a lattice
form the simplest of all quantum gases.
They can be viewed as continuous-time quantum walkers.
Quantum walks were initially introduced in the context of quantum information theory,
as a general framework to state quantum algorithms~\cite{Aharonov,Ambainis,Kempe}.
Many fundamental problems that have been studied for classical
random walks have natural counterparts in the realm of quantum walks,
often leading to results that look unexpected and surprising
from a classical point of view.
First of all, quantum walkers behave ballistically
rather than diffusively~\cite{Farhi,ToroJML}.
The survival in the presence of traps~\cite{parris,klms}
and the ballistic spreading of bound states~\cite{klmi,klmc}
provide yet other examples of qualitatively different behavior
in the classical and quantum cases.

A foremost question that has been investigated since the earlier days
of quantum mechanics concerns the decay
of individual quantum states~\cite{Gamow,condon,ww1,ww2,khalfin,winter,adler}.
In many circumstances the return probability of a quantum system to its very initial state,
also referred to as the Loschmidt echo~\cite{peres,jalabert,manfredi,gorin},
falls off exponentially in time.
Quantum revival therefore corresponds to a rare event,
somehow analogously to those described by the laws of large deviations
in classical statistical mechanics~\cite{Derrida,Touchette}.
The return probability has been recently studied
for various systems of bosons or fermions confined to one dimension.
There, a wide variety of temporal decays can be found,
ranging from a power law to a superexponential decay.
These recent works include two-particle systems~\cite{Taniguchi,Luna,Hiller},
many-particle systems~\cite{Campo1,Valle,Granot,Campo2},
and systems with infinitely many particles~\cite{Stephan2011,vsdh,Stephan2017}.
For instance, for free lattice fermions with domain-wall initial condition,
the return probability was found to obey a pure Gaussian decay
in time~\cite{Stephan2011,vsdh}.
More recently, the XXZ quantum spin chain
with domain-wall initial condition
has been investigated in the whole massless phase,
where the anisotropy parameter obeys $\abs{\Delta}<1$~\cite{Stephan2017}.
There, the decay of the return probability is found to be either Gaussian or exponential,
depending on whether the angle parameter~$\gamma$ (defined by $\Delta=\cos\gamma$) 
is commensurate to~$\pi$ or not.
This highly discontinuous asymptotic behavior is attributed to integrability.

The aim of the present work is to study the return probability
for a system of~$N$ non-interacting fermions hopping on a one-dimensional lattice.
The fermions are launched from the most compact state
where they occupy $N$ successive lattice sites.
We consider two different settings:
an infinite chain (section~\ref{r})
and a semi-infinite chain bounded by an impenetrable wall (section~\ref{w}).
In the latter situation, the compact initial state lies near the wall.
We first derive general expressions for the return probability at arbitrary finite times
(sections~\ref{rgal} and~\ref{wgal}).
We then perform an asymptotic analysis of the regime of late times
(sections~\ref{rlong} and~\ref{wlong}).
In both settings
the return probability manifests a dependence on the parity of the fermion number,
oscillating forever for odd $N$ and decaying monotonically for even $N$.
In all cases it falls off as a power of time,
whose exponent also slightly depends on the parity of $N$,
and is roughly twice smaller than what it would be in the continuum limit.
We also derive closed-form expressions for the amplitudes
of the power-law decay of the return probability in all these situations.
This derivation relies on the usage of Mehta integrals~\cite[Ch.~17]{mehta}.
Our results suggest a non-trivial crossover behavior
in the scaling regime where time $t$ and the fermion number $N$
are both large and proportional to each other
(sections~\ref{rlarge} and~\ref{wlarge}).
The determination of the corresponding scaling functions $F$ and $F\w$
remains a challenging open problem.
Section~\ref{disc} contains a brief summary and a discussion of our findings.
Two appendices are respectively devoted to the Andr\'eief identity~(\ref{appa})
and to Barnes' $G$-function~(\ref{appb}).

\section{Free fermions on the infinite chain}
\label{r}

\begin{figure}[!ht]
\begin{center}
\includegraphics[angle=-90,width=.475\linewidth]{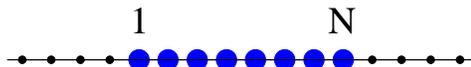}
\caption{\small
The initial configuration of the $N$ fermions on the chain.}
\label{chain}
\end{center}
\end{figure}

We consider $N$ non-interacting tight-binding fermions hopping on an infinite chain.
At the initial time ($t=0$),
the fermions are launched from $N$ consecutive sites, labeled $n=1,\dots,N$
(see~figure~\ref{chain}).

We are interested in the return probability $R_N(t)$,
i.e., the probability that the fermions occupy the same sites at a later time $t$,
and especially in the asymptotic decay of this quantity in the regime of long times.

\subsection{General expressions at finite times}
\label{rgal}

General expressions for the return probability $R_N(t)$
at arbitrary finite times can be derived as follows.
Within the formalism of the second quantization,
the Hamiltonian of the system reads
\beq
\H=\sum_n\left(a_n^\dagger a_{n+1}+a_{n+1}^\dagger a_n\right),
\eeq
in dimensionless units, with the standard fermionic anti-commutation rules
\beq
\{a_m,a_n\}=\{a_m^\dagger,a_n^\dagger\}=0,\qquad
\{a_m,a_n^\dagger\}=\delta_{m,n}.
\eeq
The many-body wavefunction $\ket{\Psi(t)}$ obeys the time-dependent Schr\"odinger equation
\beq
\ii\,\dif{\ket{\Psi}}{t}=\H\ket{\Psi},
\eeq
and so
\beq
\ket{\Psi(t)}=\e^{-\ii\H t}\ket{\Psi(0)}.
\eeq

The return probability therefore reads
\beq
R_N(t)=\abs{A_N(t)}^2,
\eeq
with
\beq
A_N(t)=\braket{\Psi(0)}{\Psi(t)}
=\braopket{\Psi(0)}{\e^{-\ii\H t}}{\Psi(0)}.
\eeq

The above expression can be made explicit
by bringing the Hamiltonian $\H$ to the diagonal form
\beq
\H=\int_0^{2\pi}\frac{\dd q}{2\pi}\,\eps_q\,\h a_q^\dagger\h a_q,
\label{hamq}
\eeq
with
\beq
\eps_q=2\cos q\qquad(0\le q\le 2\pi).
\label{epsq}
\eeq
The operators
\beq
\h a_q=\sum_n\e^{\ii nq}a_n,\qquad
\h a_q^\dagger=\sum_n\e^{-\ii nq}a_n^\dagger
\eeq
obey the anti-commutation relation
\beq
\{\h a_q,\h a_{q'}^\dagger\}=2\pi\,\delta(q-q').
\eeq

We find after some elementary algebra
\beq
A_N(t)=
\int_0^{2\pi}\etc\int_0^{2\pi}
\prod_{n=1}^N\frac{\dd q_n}{2\pi}\,\e^{-2\ii t\cos q_n}
\abs{\braket{\Psi(0)}{\q}}^2,
\label{aiam}
\eeq
where $\q=(q_1,\dots,q_N)$,
and the $q_n$ are the momenta of the $N$ fermions.
The many-body amplitude $\braket{\Psi(0)}{\q}$ is given by
a normalized Slater determinant.
In the present situation, where the fermions are launched from the sites $m=1,\dots,N$,
we have\footnote{Throughout this work, determinants are of size $N\times N$,
with indices in the range $1\le m,n\le N$.}
\beq
\braket{\Psi(0)}{\q}
=\frac{1}{\sqrt{N!}}\,\det\left(\e^{\ii mq_n}\right),
\label{amdet}
\eeq
and therefore
\beq
A_N(t)=\frac{1}{N!}
\int_0^{2\pi}\etc\int_0^{2\pi}
\prod_{n=1}^N\frac{\dd q_n}{2\pi}\,\e^{-2\ii t\cos q_n}
\abs{\det\left(\e^{\ii mq_n}\right)}^2.
\label{astart}
\eeq

The formula~(\ref{amdet}) can be recast as
\beq
\braket{\Psi(0)}{\q}
=\frac{1}{\sqrt{N!}}\prod_{n=1}^N\e^{\ii q_n}
\prod_{1\le m<n\le N}\left(\e^{\ii q_n}-\e^{\ii q_m}\right),
\label{amprod}
\eeq
where the last product is a Vandermonde determinant.
We are thus left with the following integral expression for the amplitude:
\beqa
A_N(t)=\frac{1}{N!}
&&\int_0^{2\pi}\etc\int_0^{2\pi}
\prod_{n=1}^N\frac{\dd q_n}{2\pi}\,\e^{-2\ii t\cos q_n}
\nonumber\\
&&\times\prod_{1\le m<n\le N}\abs{\e^{\ii q_n}-\e^{\ii q_m}}^2.
\label{aint}
\eeqa

Applying the Andr\'eief identity~(\ref{ai}) to~(\ref{astart}),
we get the alternative expression
\beq
A_N(t)=\det\left(\p_{m,n}(t)\right)=\det\left(J_{m-n}(2t)\right),
\label{adet}
\eeq
where
\beq
\p_{m,n}(t)=\int_0^{2\pi}\frac{\dd q}{2\pi}\,\e^{-2\ii t\cos q+\ii(n-m)q}
=\ii^{m-n}\,J_{n-m}(2t)
\label{pnt}
\eeq
is the one-body wavefunction at site $n$ at time $t$
for a particle launched from site $m$ at time 0,
and the $J_n$ are Bessel functions.

The expressions~(\ref{aint}) and~(\ref{adet}) are complementary.
The first one is more amenable to some analytical investigations,
including the asymptotic analysis of the long-time regime performed in section~\ref{rlong}.
The second one could have been obtained by more elementary means,
along the lines of~\cite{klms,klmi,klmc}.
It demonstrates explicitly that the amplitude is properly normalized,
i.e., it obeys $A_N(0)=1$.
It is also more suited for numerical evaluations and power-series expansions in $t$.

\subsection{The first few values of $N$}
\label{firstN}

It is interesting to first look at the first few values of the fermion number $N$.

\smallskip
\noindent $\bullet$ $N=1$.
For a single particle, we have $A_1(t)=J_0(2t)$, and so
\beq
R_1(t)=J_0^2(2t)\approx\frac{\cos^2(2t-\pi/4)}{\pi t}.
\eeq
The return probability therefore oscillates forever,
becoming exactly zero at infinitely many times $t_k\approx(k-1/4)\pi/2$.
These oscillations can be averaged out by replacing
the squared cosine by a factor $1/2$.
We thus obtain the power-law decay
\beq
\RR_1(t)\approx\frac{1}{2\pi t}
\eeq
for the mean return probability.

\smallskip
\noindent $\bullet$ $N=2$.
The situation of two fermions is the simplest one where quantum statistics plays a role.
We have $A_2(t)=J_0^2(2t)+J_1^2(2t)$, and so
\beq
R_2(t)=(J_0^2(2t)+J_1^2(2t))^2\approx\frac{1}{\pi^2t^2}.
\eeq
In this case the return probability falls off monotonically to zero.

\smallskip
\noindent $\bullet$ $N=3$.
We have
\beqa
A_3(t)&=&(J_0(2t)+J_2(2t))(J_0^2(2t)+2J_1^2(2t)-J_0(2t)J_2(2t))
\nonumber\\
&=&\frac{2J_1(2t)}{t}(J_0^2(2t)+J_1^2(2t))-\frac{J_0(2t)J_1^2(2t)}{t^2},
\label{p3}
\eeqa
where the first line gives the raw determinantal expression~(\ref{adet}).
The second one, obtained by means of the recursion
\beq
J_{n+1}(2t)+J_{n-1}(2t)=\frac{n}{t}J_n(2t),
\label{rec}
\eeq
is more suitable to study the late-time behavior of the return probability.
Keeping only the first group of terms in the second line of~(\ref{p3}), we obtain
\beq
R_3(t)\approx\frac{4\cos^2(2t-3\pi/4)}{\pi^3t^5}.
\eeq
The return probability again oscillates forever,
becoming exactly zero at infinitely many times $t_k\approx(k+1/4)\pi/2$.
The situation is therefore qualitatively similar to that of a single particle.
Averaging again over the oscillations, we obtain
\beq
\RR_3(t)\approx\frac{2}{\pi^3t^5}.
\eeq

\subsection{Asymptotic analysis in the long-time regime}
\label{rlong}

In this section we determine the asymptotic decay
of the return probability $R_N(t)$ in the long-time regime,
for arbitrary values of the fermion number $N$.
To do so, we evaluate the multiple integral entering~(\ref{aint})
by means of the saddle-point approximation.
Saddle points are defined by the condition that every momentum $q_n$ is either~0 or~$\pi$.
The vicinity of the most general saddle point can thus be parametrized
by choosing~$M$ momenta near 0, of the form $q_n=x_n$,
and the remaining $N-M$ momenta around~$\pi$, of the form $q_n=\pi+y_n$.
For a fixed~$M$,
there are ${N}\choose{M}$ ways of choosing which momenta are near 0 and near $\pi$.
Expanding in~(\ref{aint}) the arguments of the exponentials and the products
to quadratic order in the variables $x_n$ and $y_n$,
we obtain the estimate
\beq
A_N(t)\approx\sum_{M=0}^N\frac{I_{N,M}(t)}{M!(N-M)!},
\label{asum}
\eeq
with
\beqa
I_{N,M}(t)
&=&2^{2M(N-M)}\,\e^{2\ii(N-2M)t}
\nonumber\\
&\times&\int_{-\infty}^\infty\etc\int_{-\infty}^\infty
\prod_{n=1}^M\frac{\dd x_n}{2\pi}\,\e^{\ii t x_n^2}
\prod_{1\le m<n\le M}(x_n-x_m)^2
\\
&\times&\int_{-\infty}^\infty\etc\int_{-\infty}^\infty
\prod_{n=1}^{N-M}\frac{\dd y_n}{2\pi}\,\e^{-\ii t y_n^2}
\prod_{1\le m<n\le N-M}(y_n-y_m)^2.
\nonumber
\eeqa
The above integrals can be performed exactly.
They are indeed given be the analytical continuation to imaginary values of $a$
of the Mehta integral~\cite[Eq.~(17.6.7), $\gamma=1$]{mehta}
\beqa
\I_N(a)&=&
\int_{-\infty}^\infty\etc\int_{-\infty}^\infty
\prod_{n=1}^N\dd x_n\,\e^{-ax_n^2}
\prod_{1\le m<n\le N}(x_n-x_m)^2
\nonumber\\
&=&(2\pi)^{N/2}(2a)^{-N^2/2}\prod_{j=1}^N j!
\nonumber\\
&=&(2\pi)^{N/2}(2a)^{-N^2/2}G(N+2),
\eeqa
where $G$ is Barnes' $G$-function (see~\ref{appb}).
We thus obtain
\beqa
I_{N,M}(t)&=&2^{2M(N-M)}(2\pi)^{-N/2}\,\e^{\ii(N-2M)(2t-N\pi/4)}
\nonumber\\
&\times&G(M+2)G(N-M+2)(2t)^{M(N-M)-N^2/2}.
\eeqa

All the quantities $I_{N,M}(t)$ which enter the estimate~(\ref{asum})
thus fall off as power laws in time, albeit with an $M$-dependent exponent.
The behavior of the return probability at long times
is governed by the term $I_{N,M}(t)$ with the slowest decay.
Even and odd values of the fermion number $N$ yield different behaviors,
and have to be dealt with separately.
The emerging picture fully corroborates the observations made
in section~\ref{firstN} for the first few values of $N$.

\smallskip
\noindent $\bullet$ If $N=2m$ is even, the slowest decay is reached for $M=m$,
i.e., for equal numbers of momenta near 0 and $\pi$.
We thus obtain the estimate
\beq
A_{2m}(t)\approx\frac{2^{m(m-1)}G(m+1)^2}{\pi^m\,t^{m^2}},
\eeq
showing that the amplitude is positive and decays monotonically,
at least for large times.
The echo therefore falls off monotonically as a power law
\beq
R_{2m}(t)\approx\frac{2^{2m(m-1)}G(m+1)^4}{\pi^{2m}\,t^{2m^2}},
\label{reven}
\eeq
with exponent $2m^2=\half N^2$.

\smallskip
\noindent $\bullet$ If $N=2m+1$ is odd, the slowest decay is reached
both for $M=m$ and $M=m+1$.
We thus obtain the estimate
\beq
A_{2m+1}(t)\approx\frac{2^{m^2}G(m+1)G(m+2)}{\pi^{m+1/2}\,t^{m^2+m+1/2}}\,
\cos(2t-(2m+1)\pi/4),
\eeq
showing that the amplitude behaves for large times as an oscillatory function
modulated by a decaying power law.
Averaging over the oscillations,
we obtain the following power-law decay for the mean echo:
\beq
\RR_{2m+1}(t)\approx\frac{2^{2m^2-1}G(m+1)^2G(m+2)^2}{\pi^{2m+1}\,t^{2m^2+2m+1}},
\label{rodd}
\eeq
with exponent $2m^2+2m+1=\half(N^2+1)$.

\begin{figure}[!ht]
\begin{center}
\includegraphics[angle=-90,width=.475\linewidth]{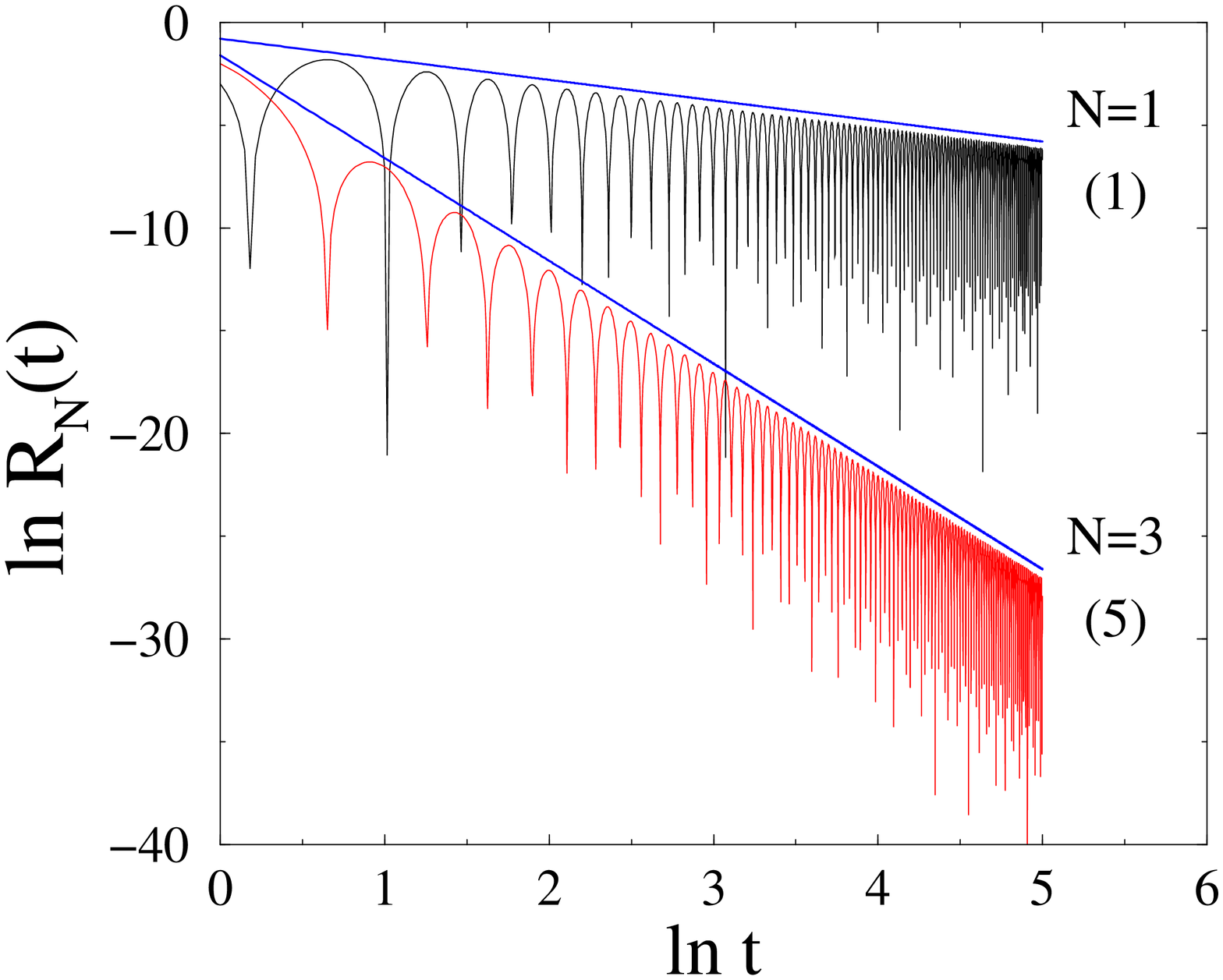}
\hskip 10pt
\includegraphics[angle=-90,width=.475\linewidth]{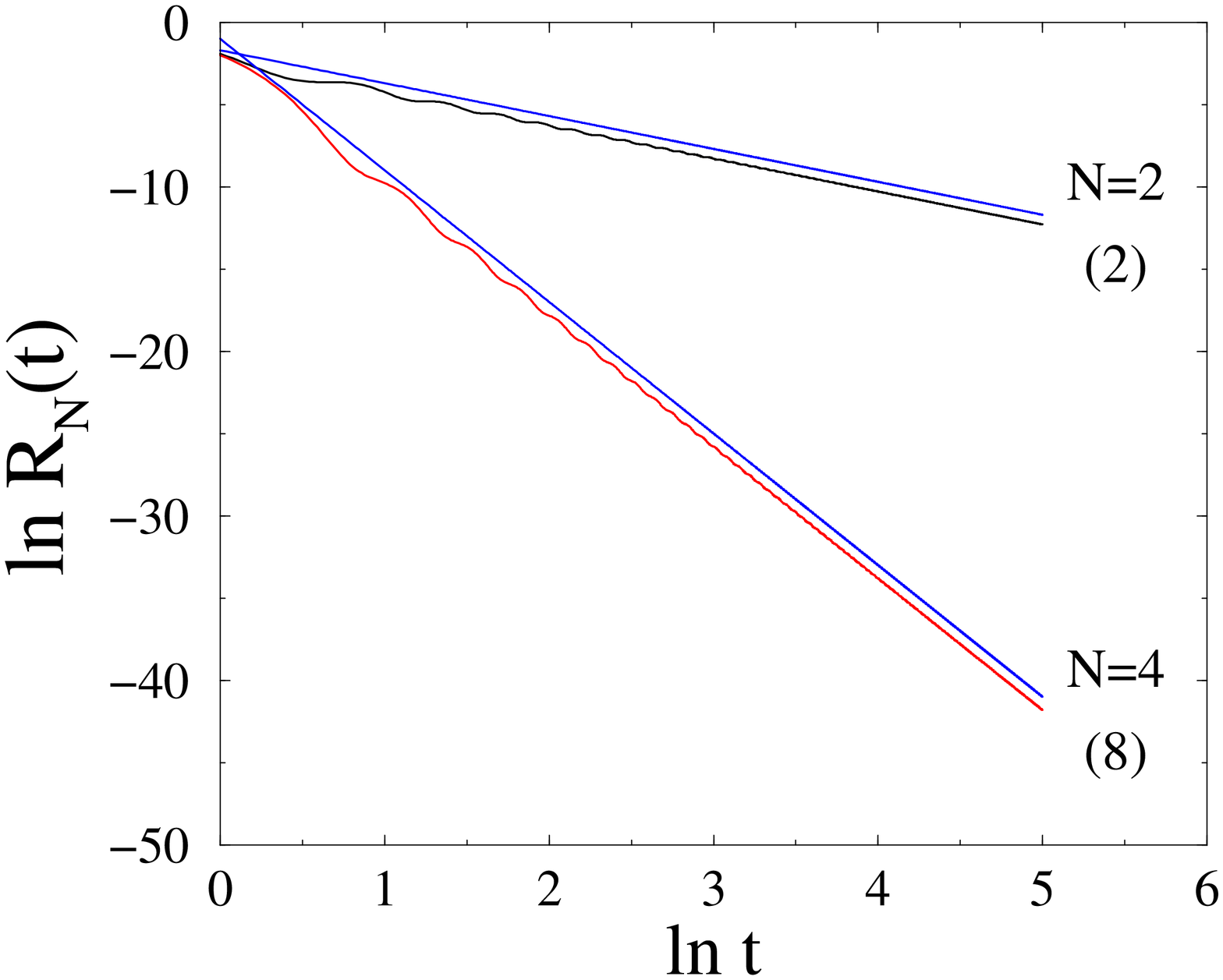}
\caption{\small
Log-log plots of the return probability $R_N(t)$ on the infinite chain
against time $t$, for $N=1$, 2, 3 and 4.
Left: odd~$N$.
Right: even~$N$.
The blue straight lines have the predicted slopes 1, 2, 5 and 8.}
\label{r1234}
\end{center}
\end{figure}

Figure~\ref{r1234} illustrates the above results, showing log-log plots
of the return probability $R_N(t)$ against time $t$ for $N=1$, 2, 3 and 4.
Data have been obtained using the expression~(\ref{adet}).
For odd $N$ (left), the echo oscillates forever.
For even~$N$ (right), it falls off monotonically
but exhibits damped oscillations,
which can be attributed to the first subleading contributions $I_{N,M}(t)$ for $N=2m$ and $M=m\pm1$.
The blue straight lines -- slightly translated for a better readability --
have the predicted slopes 1, 2, 5 and 8.

It is worth comparing the above results for lattice fermions
to the corresponding predictions in the continuum limit.
Within the present framework,
taking the continuum limit just amounts to approximating
the dispersion relation~(\ref{epsq}) by a quadratic law of the form
\beq
\eps_q\approx2-q^2,
\eeq
and therefore to restricting the saddle-point approximation
to the sector where all momenta are near zero (i.e., $M=N$).
This yields
\beq
R_N\ct(t)\approx\frac{G(N+1)^2}{2^{N(N+1)}\pi^Nt^{N^2}}.
\label{rcont}
\eeq
We still obtain a power-law decay of the return probability,
albeit with no parity effect and with a different
-- roughly twice larger -- decay exponent.

The above decay of the return probability in the continuum
can be alternatively derived by the following heuristic reasoning.
Suppose the fermions have mass $m$ and are launched from the interval $0<x<\ell$.
In reduced units ($\hbar=1$),
the typical momentum of each particle is $p\sim 1/\ell$,
and so its wavefunction spreads ballistically over a region of size
\beq
L(t)\sim\frac{t}{m\ell}.
\eeq
The return probability of one single particle therefore scales as
$R_1(t)\sim\ell/L(t)\sim m\ell^2/t$.
The exponent of this decay is in agreement with~(\ref{rodd}) and~(\ref{rcont}).
In the same setting, the modulus of the wavefunction of $N$ non-interacting fermions
at late times can be estimated as
\beq
\abs{\psi_N(x_1,\dots,x_N;t)}\sim C_N(t)\prod_{1\le m<n\le N}\abs{x_m-x_n},
\eeq
as long as all the coordinates $x_n$ are less than $L(t)$,
whereas it falls off very fast at larger distances.
Dimensional analysis implies that the normalization of the wavefunction
scales as $C_N(t)\sim L(t)^{-N^2/2}$,
and that the return probability scales as
\beq
R_N(t)\sim\left(\frac{\ell}{L(t)}\right)^{N^2}\sim\left(\frac{m\ell^2}{t}\right)^{N^2}.
\label{rheu}
\eeq
The exponent of this decay is in agreement with~(\ref{rcont}).
Furthermore, comparing the prefactors of the estimates~(\ref{rcont}) and~(\ref{rheu})
at large $N$ yields the identification $m\ell^2\approx N/(2\e^{3/2})$.
The scaling $\ell\sim\sqrt{N}$ of the length of the confining interval
is an artifact of the continuum framework.

\subsection{Large-$N$ asymptotics}
\label{rlarge}

In the $N\to\infty$ limit, the amplitude $A_N(t)$
admits the following remarkably simple expression
for all finite times:
\beq
A_\infty(t)=\lim_{N\to\infty}\det\left(J_{m-n}(2t)\right)=\e^{-t^2}.
\eeq
This expression has been derived independently in two recent works,
devoted to quantum quenches of fermionic chains~\cite{vsdh},
and to volumes of balls in unitary groups~\cite{wpct}.
The return probability therefore reads
\beq
R_\infty(t)=\e^{-2t^2}.
\label{rinf}
\eeq
It can be verified, by expanding the result~(\ref{adet})
as a power series in $t$ for the first values of $N$,
that the expressions for $R_N(t)$ and $R_\infty(t)$ start differing at order $t^{2N+2}$.
This phenomenon has already been noticed in~\cite{wpct}.

On the other hand,
for large but finite $N$,
the results of section~\ref{rlong} can be given more explicit forms.
Using the asymptotic expansion~(\ref{gbasy}),
the expressions~(\ref{reven}) for~$N$ even and~(\ref{rodd}) for $N$ odd respectively yield
\beqa
N\hbox{ even:}\quad
\ln R_N(t)=&-&\frac{N^2}{2}\left(\ln\frac{t}{N}+\frac{3}{2}\right)
\nonumber\\
&-&\frac{1}{3}\ln N+\frac{1}{3}\ln2+4\zeta'(-1)+\cdots,
\\
N\hbox{ odd:}\quad\;
\ln\RR_N(t)=&-&\frac{N^2+1}{2}\left(\ln\frac{t}{N}+\frac{3}{2}\right)
\nonumber\\
&-&\frac{1}{3}\ln N-\frac{2}{3}\ln2+\frac{3}{4}+4\zeta'(-1)+\cdots,
\eeqa
where the remainders go to zero for large $N$.
To leading order, both expressions read
\beq
\ln R_N(t)\approx-\frac{N^2}{2}\left(\ln\frac{t}{N}+\frac{3}{2}\right).
\label{nlarge}
\eeq

The results~(\ref{rinf}) and~(\ref{nlarge}) suggest
a scaling law of the form
\beq
\ln R_N(t)\approx-\frac{N^2}{2}\,F(x),\qquad x=\frac{t}{N},
\label{rsca}
\eeq
all over the regime where $t$ and $N$ are large and comparable, with
\beq
F(x)\approx\left\{\matrix{
4x^2\hfill &\quad(x\ll1),\cr
\ln x+\frad{3}{2}&\quad(x\gg1).}\right.
\label{fas}
\eeq
The form of the scaling variable $x$ reflects the ballistic nature
of the dynamics of a free tight-binding particle.

\section{Free fermions near a wall}
\label{w}

\begin{figure}[!ht]
\begin{center}
\includegraphics[angle=-90,width=.475\linewidth]{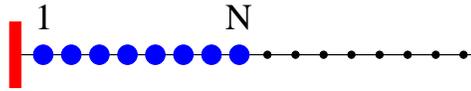}
\caption{\small
The initial configuration of the $N$ fermions near a wall
ending a semi-infinite chain.}
\label{wall}
\end{center}
\end{figure}

In this section we consider the same problem on a semi-infinite chain
ending with an impenetrable wall.
At time $t=0$, the fermions are launched from the $N$ sites
which are closest to the wall (see~figure~\ref{wall}).

We are interested in the temporal decay of the return probability $R\w_N(t)$,
where the superscript (w) reminds of the presence of a wall.

\subsection{General expressions at finite times}
\label{wgal}

General expressions for the return probability $R\w_N(t)$
at arbitrary finite times can be derived along the very lines of section~\ref{rgal}.

The presence of an impenetrable wall imposes Dirichlet boundary conditions at site $n=0$.
With these
boundary conditions, the Hamiltonian $\H$ is brought to the diagonal form
\beq
\H=\int_0^{\pi}\frac{\dd q}{\pi}\,\eps_q\,\h a_q^\dagger\h a_q,
\label{hamwq}
\eeq
with
\beq
\eps_q=2\cos q\qquad(0\le q\le\pi).
\eeq
The operators
\beq
\h a_q=\sqrt{2}\,\sum_n\sin nq\,a_n,\qquad
\h a_q^\dagger=\sqrt{2}\,\sum_n\sin nq\,a_n^\dagger
\eeq
obey the anti-commutation relation
\beq
\{\h a_q,\h a_{q'}^\dagger\}=\pi\,\delta(q-q').
\eeq

The many-body amplitude $\braket{\Psi(0)}{\q}$ is again given by
a normalized Slater determinant.
Since the fermions are launched from the sites $m=1,\dots,N$,
the latter reads
\beq
\braket{\Psi(0)}{\q}
=\sqrt{\frac{2^N}{N!}}\det\left(\sin mq_n\right).
\label{amwdet}
\eeq
The analogue of~(\ref{aiam}) therefore reads
\beq
A\w_N(t)=\frac{2^N}{N!}
\int_0^{\pi}\etc\int_0^{\pi}
\prod_{n=1}^N\frac{\dd q_n}{\pi}\,\e^{-2\ii t\cos q_n}
\left(\det\left(\sin mq_n\right)\right)^2.
\label{awstart}
\eeq

The formula~(\ref{amwdet}) can be brought to a form similar to~(\ref{amprod}).
We recall that
\beq
\sin mq=\sin q\,U_{m-1}(\cos q),
\eeq
where the $U_m$ are the Tchebyshev polynomials
of the second kind~\cite[Vol.~II, Ch.~X]{htf}.
The $m$th polynomial has degree $m$ and the leading term $U_m(z)=(2z)^m+\cdots$
We have therefore
\beqa
\det\left(\sin mq_n\right)
&=&\prod_{n=1}^N\sin q_n
\;\det\left(U_{m-1}(q_n)\right)
\nonumber\\
&=&2^{N(N-1)/2}\prod_{n=1}^N\sin q_n
\nonumber\\
&\times&\prod_{1\le m<n\le N}(\cos q_n-\cos q_m),
\eeqa
where the last product is again a Vandermonde determinant.
We are thus left with the following integral expression for the amplitude:
\beqa
A\w_N(t)=\frac{2^{N^2}}{N!}
&&\int_0^{\pi}\etc\int_0^{\pi}
\prod_{n=1}^N\frac{\dd q_n}{\pi}\,\sin^2q_n\,\e^{-2\ii t\cos q_n}
\nonumber\\
&&\times\prod_{1\le m<n\le N}(\cos q_n-\cos q_m)^2.
\label{awint}
\eeqa

Applying the Andr\'eief identity~(\ref{ai}) to~(\ref{awstart}),
we get the alternative expression
\beq
A\w_N(t)=\det\left(\p\w_{m,n}(t)\right),
\label{awdet}
\eeq
where
\beqa
\p\w_{m,n}(t)
&=&\p_{m,n}(t)-\p_{-m,n}(t)
\nonumber\\
&=&\ii^{-(n-m)}\,J_{n-m}(2t)-\ii^{-(n+m)}\,J_{n+m}(2t)
\label{pwnt}
\eeqa
is the one-body wavefunction at site $n$ at time $t$
for a particle located
on the semi-infinite chain and launched from site $m$ at time 0.
The expression~(\ref{pwnt}) can be recovered by applying the reflection principle
(i.e., the method of images) to~(\ref{pnt}).

The complementary expressions~(\ref{awint}) and~(\ref{awdet})
are the exact analogues on the semi-infinite chain
of their counterparts~(\ref{aint}) and~(\ref{adet}) on the infinite chain.

\subsection{Asymptotic analysis in the long-time regime}
\label{wlong}

In this section we determine the asymptotic decay
of the return probability $R\w_N(t)$.
We again apply the saddle-point approximation
to the multiple integral in~(\ref{awint}).
Saddle points are still defined by the condition
that every momentum $q_n$ is either~0 or~$\pi$.
The vicinity of the most general saddle point will be parametrized
by choosing~$M$ momenta near 0, of the form $q_n=x_n$,
and the remaining $N-M$ momenta around~$\pi$, of the form $q_n=\pi-y_n$.
At variance with the previous situation (section~\ref{rlong}),
the variables $x_n$ and $y_n$ are now positive.
We thus obtain the estimate
\beq
A\w_N(t)\approx2^{N^2}\sum_{M=0}^N\frac{I\w_{N,M}(t)}{M!(N-M)!},
\label{awsum}
\eeq
with
\beqa
I\w_{N,M}(t)
&=&2^{N-(N-2M)^2}\,\e^{2\ii(N-2M)t}
\nonumber\\
&\times&\int_0^\infty\etc\int_0^\infty
\prod_{n=1}^M\frac{\dd x_n}{\pi}\,x_n^2\,\e^{\ii t x_n^2}
\prod_{1\le m<n\le M}(x_n^2-x_m^2)^2
\\
&\times&\int_0^\infty\etc\int_0^\infty
\prod_{n=1}^{N-M}\frac{\dd y_n}{\pi}\,y_n^2\,\e^{-\ii t y_n^2}
\prod_{1\le m<n\le N-M}(y_n^2-y_m^2)^2.
\nonumber
\eeqa
The above integrals can still be performed exactly.
They are indeed proportional to the analytical continuation to imaginary values of $a$
of the Mehta integral~\cite[Eq.~(17.6.6), $\alpha=3/2$, $\gamma=1$]{mehta}
\beqa
\I\w_N(a)&=&
\int_{-\infty}^\infty\etc\int_{-\infty}^\infty
\prod_{n=1}^N\dd x_n\,x_n^2\,\e^{-ax_n^2}
\prod_{1\le m<n\le N}(x_n^2-x_m^2)^2
\nonumber\\
&=&a^{-N(2N+1)/2}\prod_{j=1}^Nj!\,\Gamma(j+1/2)
\nonumber\\
&=&\pi^{N/2}(2a)^{-N(2N+1)/2}\sqrt{N!\,G(2N+2)},
\eeqa
where $G$ is again Barnes' $G$-function (see~\ref{appb}).
We thus obtain
\beqa
I\w_{N,M}(t)&=&2^{-(N-2M)^2}\pi^{-N/2}\,\e^{\ii(N-2M)(2t-(N+1/2)\pi/4)}
\nonumber\\
&\times&\sqrt{M!(N-M)!\,G(2M+2)G(2N-2M+2)}
\nonumber\\
&\times&(2t)^{2M(N-M)-N(2N+1)/2}.
\eeqa

The behavior of the return probability at long times
is still governed by the term with the slowest decay.
Even and odd values of the fermion number $N$ again yield
different behaviors.

\smallskip
\noindent $\bullet$ If $N=2m$ is even, the slowest decay is reached for $M=m$,
i.e., for equal numbers of momenta near 0 and $\pi$.
We thus obtain the estimate
\beq
A\w_{2m}(t)\approx\frac{2^{m(2m-1)}G(2m+2)}{\pi^m\,m!\,t^{m(2m+1)}},
\eeq
showing that the amplitude is positive and decays monotonically,
at least for large times.
The echo therefore falls off monotonically as a power law
\beq
R\w_{2m}(t)\approx\frac{2^{2m(2m-1)}G(2m+2)^2}{\pi^{2m}\,m!^2\,t^{2m(2m+1)}},
\label{rweven}
\eeq
with exponent $2m(2m+1)=N(N+1)$.

\smallskip
\noindent $\bullet$ If $N=2m+1$ is odd, the slowest decay is reached
both for $M=m$ and $M=m+1$.
We thus obtain the estimate
\beq
A\w_{2m+1}(t)\approx\frac{2^{m(2m+1)}G(2m+3)}{\pi^{m+1/2}\,m!\,t^{2m^2+3m+3/2}}\,
\cos(2t-(2m+3/2)\pi/4),
\eeq
showing that the amplitude behaves for large times as an oscillatory function
modulated by a decaying power law.
Averaging over the oscillations,
we obtain the following power-law decay for the mean echo:
\beq
\RR\w_{2m+1}(t)\approx\frac{2^{2m(2m+1)-1}G(2m+3)^2}{\pi^{2m+1}\,m!^2\,t^{4m^2+6m+3}},
\label{rwodd}
\eeq
with exponent $4m^2+6m+3=N^2+N+1$.

\begin{figure}[!ht]
\begin{center}
\includegraphics[angle=-90,width=.475\linewidth]{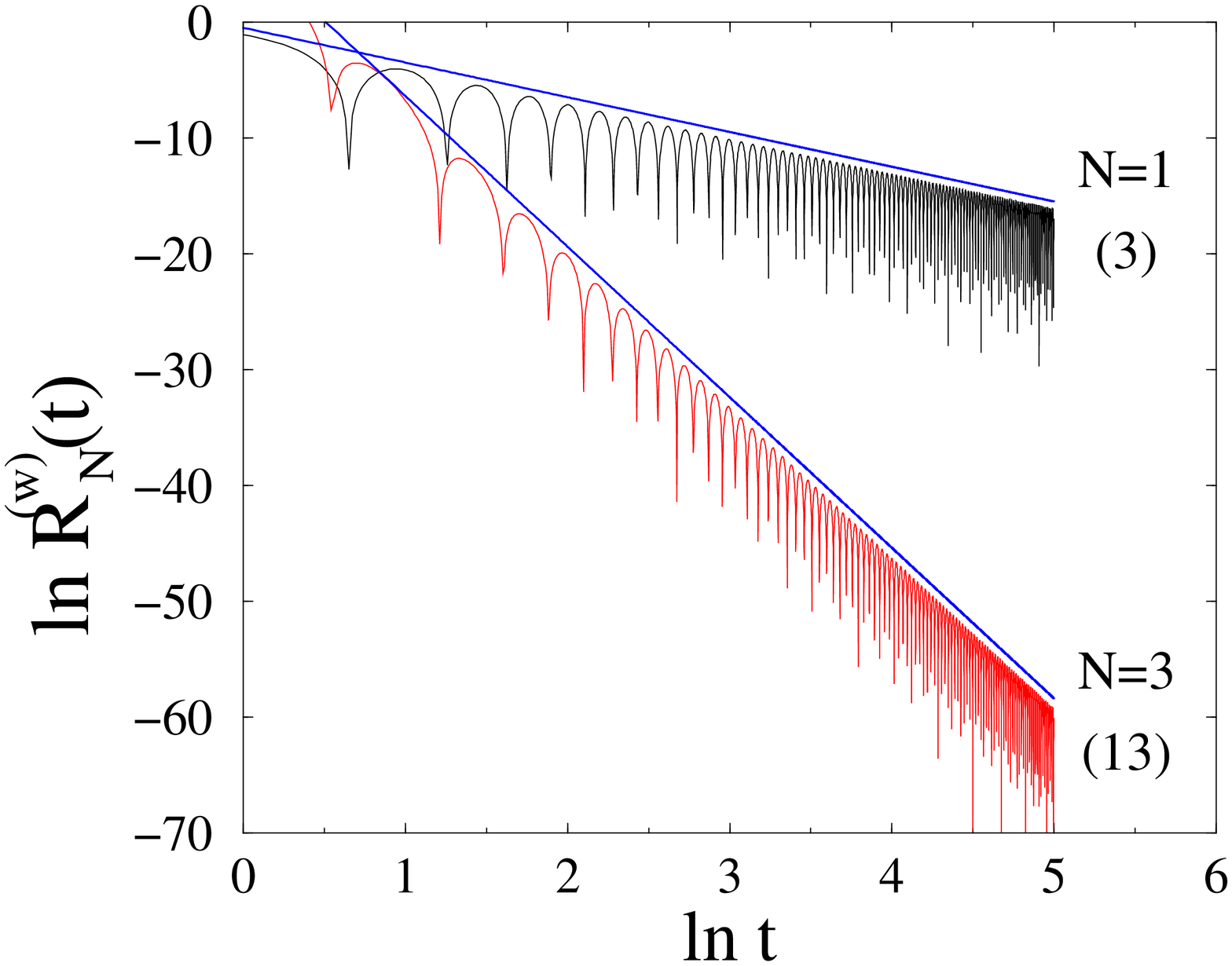}
\hskip 10pt
\includegraphics[angle=-90,width=.475\linewidth]{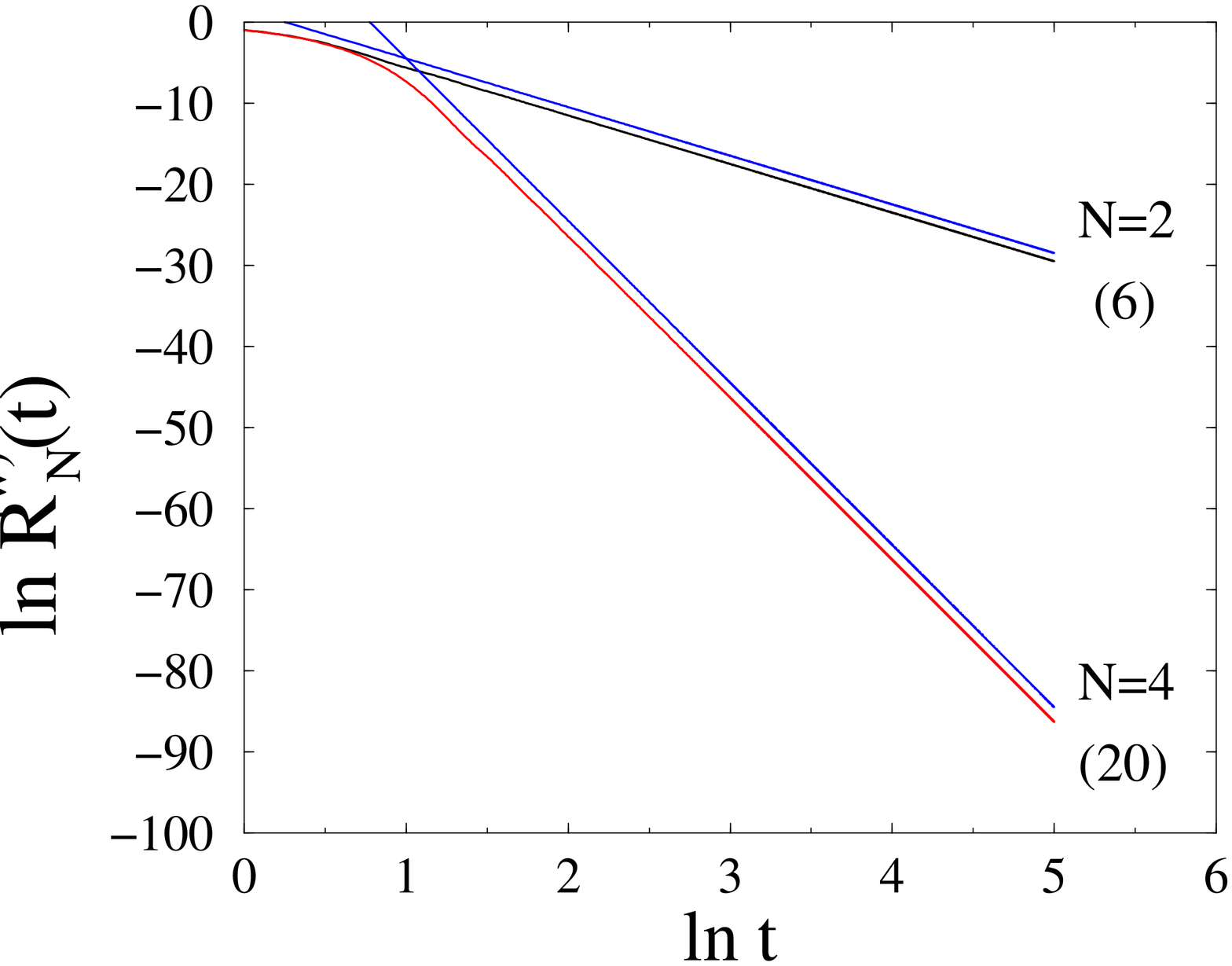}
\caption{\small
Log-log plots of the return probability $R\w_N(t)$ near a wall
ending a semi-infinite chain against time $t$, for $N$ up to 4.
Left: odd~$N$.
Right: even~$N$.
The blue straight lines have the predicted slopes 3, 6, 13 and 20.}
\label{w1234}
\end{center}
\end{figure}

Figure~\ref{w1234} illustrates the above results, showing log-log plots
of the return probability $R\w_N(t)$
against time $t$ for $N=1$, 2, 3 and 4.
Data have been obtained using the expression~(\ref{awdet}).
For odd $N$ (left), the echo oscillates forever.
For even~$N$ (right), it falls off monotonically
but exhibits damped oscillations,
which can again be attributed
to the first subleading contributions $I\w_{N,M}(t)$ for $N=2m$ and $M=m\pm1$.
The blue straight lines -- slightly translated for a better readability --
have the predicted slopes 3, 6, 13 and 20.

It is again worth comparing the above results for lattice fermions
to the corresponding prediction in the continuum limit.
Restricting the saddle-point approximation
to the sector where all momenta are near zero (i.e., $M=N$) yields
\beq
R_N\wct(t)\approx\frac{G(2N+2)}{2^{N(2N+1)}\pi^NN!\,t^{N(2N+1)}}.
\label{wcont}
\eeq
We again obtain a power-law decay of the return probability,
albeit with no parity effect and with a different, larger decay exponent.

The above power-law decay can still be recovered by means of heuristic reasoning.
Suppose the fermions are launched from the interval $0<x<\ell$ near the wall.
The modulus of the many-body wavefunction at time $t$ now reads approximately
\beq
\abs{\psi_N\w(x_1,\dots,x_N;t)}\sim C_N\w(t)
\prod_{n=1}^N x_n\prod_{1\le m<n\le N}\abs{x_m^2-x_n^2}.
\eeq
Dimensional analysis determines the scaling of
the normalization of the wavefunction,
$C_N\w(t)\sim L(t)^{-N(2N+1)}$,
and of the return probability,
\beq
R_N\w(t)\sim\left(\frac{\ell}{L(t)}\right)^{N(2N+1)}
\sim\left(\frac{m\ell^2}{t}\right)^{N(2N+1)}.
\label{wheu}
\eeq
A very similar result can be found in~\cite{Campo1}.
The decay exponent of the above estimate agrees with~(\ref{wcont}).
Furthermore, comparing the prefactors of the estimates~(\ref{wcont}) and~(\ref{wheu})
at large $N$ yields $m\ell^2\approx N/\e^{3/2}$.
The latter estimate is twice larger than its counterpart on the infinite line,
given below~(\ref{rheu}).
An intuitive interpretation of this factor two will be given below~(\ref{wsca}).

\subsection{Large-$N$ asymptotics}
\label{wlarge}

In the $N\to\infty$ limit,
the return probability is expected to be equal
to the square root of its counterpart~(\ref{rinf}) on the infinite chain, namely
\beq
R\w_\infty(t)=\e^{-t^2}.
\label{rwinf}
\eeq
A heuristic way of showing this goes as follows.
On the infinite chain, the compact fermionic state has two ends,
and can therefore decay through either end,
whereas it has only one right end if confined near a wall.
Hence we can expect $R_\infty(t)=(R\w_\infty(t))^2$.
The latter result can be checked by expanding the expression~(\ref{awdet})
as a power series in $t$ for the first values of $N$.
Doing so confirms our expectation
and shows that $R\w_N(t)$ and $R\w_\infty(t)$ again start differing at order $t^{2N+2}$.

On the other hand,
using the asymptotic formula~(\ref{gbasy}),
we can still derive more explicit forms of the above results for large $N$.
The expressions~(\ref{rweven}) for $N$ even and~(\ref{rwodd}) for $N$ odd
respectively yield
\beqa
N\hbox{ even:}\quad
\ln R\w_N(t)=&-&N(N+1)\left(\ln\frac{t}{2N+1}+\frac{3}{2}\right)
\nonumber\\
&-&\frac{1}{6}\ln N+\ln 2-\frac{3}{8}+2\zeta'(-1)+\cdots,
\\
N\hbox{ odd:}\quad\;
\ln\RR\w_N(t)=&-&(N^2+N+1)\left(\ln\frac{t}{2N+1}+\frac{3}{2}\right)
\nonumber\\
&-&\frac{1}{6}\ln N-2\ln 2+\frac{9}{8}+2\zeta'(-1)+\cdots
\eeqa
The argument $t/(2N+1)$ of the logarithms
has been chosen in order to minimize the order of magnitude of the correction terms
given in the second lines of the above expressions.
To leading order, both results read
\beq
\ln R\w_N(t)\approx-N^2\left(\ln\frac{t}{2N}+\frac{3}{2}\right).
\label{wnlarge}
\eeq

The results~(\ref{rwinf}) and~(\ref{wnlarge}) again suggest a scaling law of the form
\beq
\ln R\w_N(t)\approx-N^2\,F\w(x),\qquad x=\frac{t}{2N}.
\label{wsca}
\eeq
The effective fermion number $2N$ entering the scaling variable $x$
can be interpreted as the total distance
to be traveled by an excitation entering the compact fermionic state from the right,
bouncing at the wall, and exiting from the right.
Finally, the scaling function $F\w(x)$
obeys the very same asymptotics~(\ref{fas}) as $F(x)$ both for $x\ll1$ and for $x\gg1$.
This observation might be more than a coincidence,
in the sense that both scaling functions might be identical.

\section{Discussion}
\label{disc}

We studied the quantum return probability for a system of $N$ non-interacting
lattice fermions launched from $N$ consecutive sites,
either on the infinite chain or near an impenetrable wall ending a semi-infinite chain.

In each case we derived exact expressions for the return probability valid for all fermion numbers $N$ and time $t$.
We thus obtained two complementary kinds of expressions,
namely integral formulas (see~(\ref{aint}) and~(\ref{awint})),
which are the natural outcome of the second-quantized formalism,
and determinantal formulas (see~(\ref{adet}) and~(\ref{awdet})),
which could have been obtained by a more elementary, first-quantized approach as well.
We deduced the asymptotic long-time behavior of the return probability
by evaluating the integral formulas by means of the saddle-point method.
Even and odd values of the fermion number $N$ yield different qualitative behaviors,
as well as slightly different expressions for the decay exponents.
For even~$N$, the echo falls off monotonically as a power law.
For odd~$N$, it exhibits periodic oscillations modulated by a decaying power law.
This qualitative dependence on the parity of the fermion number is a pure lattice effect,
which is absent in the continuum limit.
The return probability of $N$ particles thus provides
yet another example of a situation where quantum dynamics
exhibits qualitatively different features on the lattice and in the continuum.
The exponents characterizing the temporal decay of the return probability are gathered in table~\ref{expos}.

\begin{table}[!ht]
\begin{center}
\begin{tabular}{|c|c||c|c|}
\hline
infinite chain & infinite chain & near a wall & near a wall\\
(lattice) & (continuum) & (lattice) & (continuum)\\
\hline
$\!\left\{
\matrix{
N\hbox{~even:}\hfill &\half{N^2}^\vv\hfill\cr
N\hbox{~odd:}\hfill &\half(N^2+1)_\vv\hfill\cr
}\right.$
& $N^2$
&
$\!\left\{
\matrix{
N\hbox{~even:}\hfill & N(N+1)\hfill\cr
N\hbox{~odd:}\hfill & N^2+N+1\hfill\cr
}\right.$
& $N(2N+1)$\\
\hline
\end{tabular}
\caption{Exponents of the temporal decay of the return probability
on the infinite chain and near a wall.
Comparison between the values of the exponents for lattice fermions
and the predictions of the continuum limit.}
\label{expos}
\end{center}
\end{table}

Our results also yield explicit expressions for the prefactors
of the asymptotic power-law decay of the return probability,
in both geometries and for all values of the fermion number $N$.
The key point in the derivation of these results
has been the use of Mehta integrals,
which have been extensively used in random matrix theory~\cite[Ch.~17]{mehta}
and can be viewed as limiting cases of the Selberg integral.
Reference~\cite{selberg} provides a historical overview
of the Selberg and related integrals,
whereas the recent work~\cite{PRAMajumdar} and the references therein
mention yet other connections between random matrix theory and systems of free fermions.
Our expressions for the prefactors of the power-law decays
involve as an essential ingredient Barnes' $G$-function,
which is also ubiquitous in random matrix theory.
Table~\ref{asy} gives our predictions in factorized form up to $N=10$.

\begin{table}[!ht]
\begin{center}
\begin{tabular}{|c|c|c||c|c|c|}
\hline
& infinite chain & near a wall & & infinite chain & near a wall\\
$N$ odd & $\RR_N(t)$ & $\RR\w_N(t)$ &
$N$ even & $R_N(t)$ & $R\w_N(t)_\vv$\\
& (\ref{rodd}) & (\ref{rwodd}) &
& (\ref{reven})$^\vv$ & (\ref{rweven})\\
\hline
1 & $\frad{1}{2\pi t}$ & $\frad{1}{2\pi t^3}$ &
2 & $\frad{1}{\pi^2 t^2_\vv}$ & $\frad{2^4}{\pi^2 t^6}$\\
3 & $\frad{2}{\pi^3 t^5}$ & $\frad{2^9 3^2}{\pi^3 t^{13}}$ &
4 & $\frad{2^4}{\pi^4 t^8_\vv}$ & $\frad{2^{20} 3^4}{\pi^4 t^{20}}$\\
5 & $\frad{2^9}{\pi^5 t^{13}}$ & $\frad{2^{33} 3^6 5^2}{\pi^5 t^{31}}$ &
6 & $\frad{2^{16}}{\pi^6 t^{18}_\vv}$ & $\frad{2^{52} 3^8 5^4}{\pi^6 t^{42}}$\\
7 & $\frad{2^{23} 3^2}{\pi^7 t^{25}}$ & $\frad{2^{71} 3^{12} 5^6 7^2}{\pi^7 t^{57}}$ &
8 & $\frad{2^{32} 3^4}{\pi^8 t^{32}_\vv}$ & $\frad{2^{96} 3^{16} 5^8 7^4}{\pi^8 t^{72}}$\\
9 & $\frad{2^{45} 3^6}{\pi^9 t^{41}}$ & $\frad{2^{125} 3^{24} 5^{10} 7^6}{\pi^9 t^{91}}$ &
10 & $\frad{2^{60} 3^8}{\pi^{10} t^{50}_\vv}$ & $\frad{2^{160} 3^{32} 5^{12} 7^8}{\pi^{10} t^{110}}$\\
\hline
\end{tabular}
\caption{Asymptotic temporal decay of the return probability
on the infinite chain and near a wall, for fermion numbers up to $N=10$.
The exact prefactors are given in factorized form.}
\label{asy}
\end{center}
\end{table}

The behavior of our results at large fermion numbers
led us to hypothesize that the return probability
obeys the scaling laws~(\ref{rsca}) and~(\ref{wsca}),
with arguments $x=t/N$ and $x=t/(2N)$,
in the regime where~$t$ and $N$ are both large and comparable.
The form of these scaling variables reflects the ballistic nature
of the motion of a free tight-binding particle.
It would clearly be desirable to investigate this ballistic scaling regime
in a more thorough fashion,
in order to eventually determine the analytic form of the scaling functions $F$ and $F\w$,
and thus prove or disprove their conjectural equality suggested by the similarity
of their behavior at small and large values of their argument.
Such an investigation would certainly imply the usage of more advanced techniques.
There is indeed no simple way to generalize the saddle-point approach
used here in order to encompass the whole ballistic scaling regime.

The framework of the present study could also be extended in several other directions.
One could investigate the return probability of $N$ fermions
launched from other types of localized states,
either on the chain,
such as compactly supported but not most compact states,
or in the continuum,
with arbitrary initial states whose many-body wavefunction is localized
in the center-of-mass coordinate.
A further direction of research consists in replacing free fermions
by other integrable systems of interacting particles in one dimension.
Finally, higher-dimensional settings are also worth being envisaged.
We intend to return to some of these matters in the future.

\ack

It is a pleasure to thank Jean-Paul Blaizot, Adolfo del Campo, Bertrand Eynard
and Jean-Marie St\'ephan for fruitful discussions.

\appendix

\section{The Andr\'eief identity}
\label{appa}

The Andr\'eief identity
\beqa
&&\int_a^b\etc\int_a^b\prod_{n=1}^N\rho(x_n)\,\dd x_n
\det\left(f_m(x_n)\right)\;\det\left(g_m(x_n)\right)
\nonumber\\
&&=N!\,\det\biggl(\int_a^b\rho(x)\,\dd x\,f_m(x)g_n(x)\biggr)
\label{ai}
\eeqa
relates the integral of the product of the determinants
built upon two families of functions to the determinant of their scalar products.
This identity comes in many guises in various branches of the mathematical literature,
concerning especially orthogonal polynomials and random matrix theory.
Although it seems to appear for the first time in print in an article
by Andr\'eief in 1883~\cite{andrei},
it is also associated with other names, including Cauchy-Binet, Gram and Heine.
Here, it allows us to respectively derive~(\ref{adet}) and~(\ref{awdet})
from~(\ref{astart}) and~(\ref{awstart}).

Let us give an elementary proof of the above identity for the sake of completeness.
Starting from the left-hand side,
let us introduce the Leibniz expansions of the determinants:
\beqa
\det\left(f_m(x_n)\right)&=&\sum_\sigma\sgn\sigma\prod_{n=1}^Nf_{\sigma_n}(x_n),
\nonumber\\
\det\left(g_m(x_n)\right)&=&\sum_\tau\sgn\tau\prod_{n=1}^Ng_{\tau_n}(x_n),
\eeqa
where $\sigma$ and $\tau$ are permutations acting on $N$ symbols,
and $\sgn\sigma=\pm1$ and $\sgn\tau=\pm1$ are their signatures.
We thus obtain
\beq
I_N=\sum_{\sigma,\tau}\sgn\sigma\sgn\tau
\int_a^b\etc\int_a^b
\prod_{n=1}^N\rho(x_n)\,\dd x_n\,f_{\sigma_n}(x_n)g_{\tau_n}(x_n).
\eeq
The integrand is now a product, and so
\beq
I_N=\sum_{\sigma,\tau}\sgn\sigma\sgn\tau
\prod_{n=1}^N\int_a^b\rho(x)\,\dd x\,f_{\sigma_n}(x)g_{\tau_n}(x).
\eeq
For fixed permutations $\sigma$ and $\tau$,
let us change the index from $n$ to $m=\tau_n$.
We have then $\sigma_n=\mu_m$, where $\mu=\sigma\cdot\tau^{-1}$,
and so $\sgn\mu=\sgn\sigma\sgn\tau$.
For fixed $\tau$, the sum over $\sigma$ can be replaced by a sum over $\mu$.
The sum over $\tau$ simply yields a factor $N!$.
We thus obtain
\beq
I_N=N!\sum_\mu\sgn\mu
\prod_{n=1}^N\int_a^b\rho(x)\,\dd x\,f_{\mu_m}(x)g_m(x).
\eeq
The sum over $\mu$ is nothing but the Leibniz expansion of the determinant
given in the right-hand side of~(\ref{ai}).

On the infinite chain, the expression~(\ref{astart}) of the amplitude $A_N(t)$
is proportional to the left-hand side of~(\ref{ai}), with
\beqa
x_n=q_n,\qquad
\rho(x_n)=\e^{-2\ii t\cos q_n},
\nonumber\\
f_m(x_n)=\e^{\ii mq_n},\qquad
g_m(x_n)=\e^{-\ii mq_n}.
\eeqa
Applying the identity~(\ref{ai}) yields~(\ref{adet}).

On the semi-infinite chain, the expression~(\ref{awstart}) of the amplitude $A\w_N(t)$
is proportional to the left-hand side of~(\ref{ai}), with the same $\rho(x_n)$ and
\beq
f_m(x_n)=g_m(x_n)=\sqrt{2}\,\sin mq_n.
\eeq
Applying the identity~(\ref{ai}) yields~(\ref{awdet}).

\section{Barnes' $G$-function}
\label{appb}

Barnes' $G$-function shares many common features with Euler's $\Gamma$-function.
This appendix summarizes the main properties of both functions,
which can be found in the Wikipedia article~\cite{wbarnes}
or in the {\it Digital Library of Mathematical Functions}~\cite[Ch.~5.17]{nist}.

Euler's $\Gamma$-function and Barnes' $G$-function
are meromorphic functions in the complex plane
obeying the recursion relations
\beq
\Gamma(z+1)=z\Gamma(z),\qquad G(z+1)=\Gamma(z)G(z),
\eeq
with appropriate regularity conditions.

When $z$ is a positive integer,
Euler's $\Gamma$-function becomes the usual factorial:
\beq
\Gamma(n+1)=n!,
\eeq
whereas Barnes' $G$-function becomes the `superfactorial':
\beq
G(n+2)=\prod_{k=1}^n k!=\prod_{\ell=1}^n\ell^{n+1-\ell}=\prod_{1\le i<j\le n+1}(j-i).
\eeq
We have in particular $\Gamma(1)=\Gamma(2)=1$ and $G(1)=G(2)=G(3)=1$.
The `super\-factorial' numbers $G(n+2)$ appear in the OEIS~\cite{oeis}
as sequence number A000178,
together with many further properties and references.

Euler's $\Gamma$-function and Barnes' $G$-function
have the following asymptotic expan\-sions as $z\to+\infty$:
\beqa
\ln\Gamma(z+1)
&=&\left(z+\frac{1}{2}\right)\ln z-z+\ln\!\sqrt{2\pi}+\cdots,
\\
\ln G(z+1)
&=&\left(\frac{z^2}{2}-\frac{1}{12}\right)\ln z-\frac{3z^2}{4}
+z\ln\!\sqrt{2\pi}+\zeta'(-1)+\cdots,
\label{gbasy}
\eeqa
where $\zeta'(-1)=-0.165\,421\,143\dots$ ($\zeta$ being Riemann's $\zeta$-function),
and the remainders go to zero for large $z$.


\section*{References}

\begin{thebibliography}{99}

\bibitem{Dalibard} Bloch I, Dalibard J and Nascimb\`ene S 2012
Quantum simulations with ultracold quantum gases
{\it Nature Phys.} {\bf 8} 267

\bibitem{Paredes} Paredes B, Widera A, Murg V, Mandel O, F\"olling S, Cirac I, Shlyapnikov G V, H\"ansch T W and Bloch I 2004
Tonks-Girardeau gas of ultracold atoms in an optical lattice
{\it Nature} {\bf 429} 277

\bibitem{Kinoshita} Kinoshita T, Wenger T and Weiss D S 2006
A quantum Newton's cradle
{\it Nature} {\bf 440} 900

\bibitem{Cazalilla} Cazalilla M A and Rigol M 2011
Focus on dynamics and thermalization in isolated quantum many-body systems
{\it New J. Phys.} {\bf 12} 055006

\bibitem{Polkovnikov} Polkovnikov A, Sengupta K, Silva A and Vengalatorre M 2011
Nonequilibrium dynamics of closed interacting quantum systems
{\it Rev. Mod. Phys.} {\bf 83} 863

\bibitem{Eisert} Eisert J, Friesdorf M and Gogolin C 2015
Quantum many-body systems out of equilibrium
{\it Nature Phys.} {\bf 11} 124

\bibitem{Haroche} Haroche S and Raimond J M 2006
{\it Exploring the Quantum: Atoms, Cavities, and Photons}
(Oxford: Oxford University Press)

\bibitem{Aharonov} Aharonov Y, Davidovich L and Zagury N 1993
Quantum random walks
{\it Phys. Rev. A} {\bf 48} 1687

\bibitem{Ambainis} Ambainis A 2003
Quantum walks and their algorithmic applications
{\it Int. J. Quant. Inf.} {\bf 1} 507

\bibitem{Kempe} Kempe J 2003
Quantum random walks: An introductory overview
{\it Contemp. Phys.} {\bf 44} 307

\bibitem{Farhi} Farhi E and Gutmann S 1998
Quantum computation and decision trees
{\it Phys. Rev. A} {\bf 58} 915

\bibitem{ToroJML} de Toro Arias S and Luck J M 1998
Anomalous dynamical scaling and bifractality in the one-dimensional Anderson model
{\it J. Phys. A: Math. Gen.} {\bf 31} 7699

\bibitem{parris} Parris P E 1989
One-dimensional trapping kinetics at zero temperature
{\it Phys. Rev. Lett.} {\bf 62} 1392

\bibitem{klms} Krapivsky P L, Luck J M and Mallick K 2014
Survival of classical and quantum particles in the presence of traps
{\it J. Stat. Phys.} {\bf 154} 1430

\bibitem{klmi} Krapivsky P L, Luck J M and Mallick K 2015
Interacting quantum walkers: Two-body bosonic and fermionic bound states
{\it J. Phys. A: Math. Theor.} {\bf 48} 475301

\bibitem{klmc} Krapivsky P L, Luck J M and Mallick K 2016
Quantum centipedes: Collective dynamics of interacting quantum walkers
{\it J. Phys. A: Math. Theor.} {\bf 49} 335303

\bibitem{Gamow} Gamow G 1928
Zur Quantentheorie des Atomkernes
{\it Z. Phys.} {\bf 51} 204

\bibitem{condon} Gurney R W and Condon E U 1929
Quantum mechanics and radioactive disintegration
{\it Phys. Rev.} {\bf 33} 127

\bibitem{ww1} Weisskopf V and Wigner E 1930
Berechnung der nat\"urlichen Linienbreite auf Grund der Dirac\-schen Lichttheorie
{\it Z. Phys.} {\bf 63} 54

\bibitem{ww2} Weisskopf V and Wigner E 1930
\"Uber die nat\"urliche Linienbreite in der Strahlung des harmo\-nischen Oszillators
{\it Z. Phys.} {\bf 65} 18

\bibitem{khalfin} Khalfin L A 1958
Contribution to the decay theory of a quasi-stationary state
{\it Sov. Phys. JETP} {\bf 6} 1053 [Zh. Eks. Teor. Fiz. {\bf 33} 1371 (1957)]

\bibitem{winter} Winter R G 1961
Evolution of a quasi-stationary state
{\it Phys. Rev.} {\bf 123} 1503

\bibitem{adler} Adler S L 2003
Weisskopf-Wigner decay theory for the energy-driven stochastic Schr\"{o}dinger equation
{\it Phys. Rev. D} {\bf 67} 025007

\bibitem{peres} Peres A 1984
Stability of quantum motion in chaotic and regular systems
{\it Phys. Rev. A} {\bf 30} 1610

\bibitem{jalabert} Jalabert R A and Pastawski H M 2001
Environment-independent decoherence rate in classically chaotic systems
{\it Phys. Rev. Lett.} {\bf 86} 2490

\bibitem{manfredi} Manfredi G and Hervieux P A 2006
Loschmidt echo in a system of interacting electrons
{\it Phys. Rev. Lett.} {\bf 97} 190404

\bibitem{gorin} Gorin T, Prosen T, Seligman T H and Znidaric M 2006
Dynamics of Loschmidt echoes and fidelity decay
{\it Phys. Rep.} {\bf 435} 33

\bibitem{Derrida} Derrida B 2007
Non-equilibrium steady states: Fluctuations and large deviations of the density and of the current
{\it J. Stat. Mech.} P07023

\bibitem{Touchette} Touchette H 2009
The large deviation approach to statistical mechanics
{\it Phys. Rep.} {\bf 478} 1

\bibitem{Taniguchi} Taniguchi T and Sawada S I 2011
Escape behavior of quantum two-particle systems with Coulomb interactions
{\it Phys. Rev. E} {\bf 83} 026208

\bibitem{Luna} Garc\'{i}a-Calder\'{o}n G and Mendoza-Luna L G 2011
Time evolution of decay of two identical quantum particles
{\it Phys. Rev. A} {\bf 84} 032106

\bibitem{Hiller} Hunn S, Zimmermann K, Hiller M and Buchleitner A 2013
Tunneling decay of two interacting bosons in an asymmetric double-well potential: A spectral approach
{\it Phys. Rev. A} {\bf 87} 043626

\bibitem{Campo1} del Campo A 2011
Long-time behavior of many-particle quantum decay
{\it Phys. Rev. A} {\bf 84} 012113

\bibitem{Valle} Longhi S and della Valle G 2012
Many-particle quantum decay and trapping: The role of statistics and Fano resonances
{\it Phys. Rev. A} {\bf 86} 012112

\bibitem{Granot} Marchewka A and Granot E 2015
Role of quantum statistics in multi-particle decay dynamics
{\it Ann. Phys.} {\bf 355} 348

\bibitem{Campo2} del Campo A 2016
Exact quantum decay of an interacting many-particle system: The Calogero-Sutherland model
{\it New J. Phys.} {\bf 18} 015014

\bibitem{Stephan2011} St\'ephan J M and Dubail J 2011
Local quantum quenches in critical one-dimensional systems: Entanglement, the Loschmidt echo, and light-cone effects
{\it J. Stat. Mech.} P08019

\bibitem{vsdh} Viti J, St\'ephan J M, Dubail J and Haque M 2016
Inhomogeneous quenches in a free fermionic chain: Exact results
{\it EPL} {\bf 115} 40011

\bibitem{Stephan2017} St\'ephan J M 2017
Return probability after a quench from a domain-wall initial state in the spin-1/2 XXZ chain
{\it J. Stat. Mech.} P103108

\bibitem{mehta} Mehta M 1990
{\it Random Matrices} 2nd ed
(London: Academic)

\bibitem{wpct} Wei L, Pitaval R A, Corander J and Tirkkonen O 2017
From random matrix theory to coding theory: Volume of a metric ball in unitary groups
{\it IEEE Trans. Information Theory} {\bf 63} 2814

\bibitem{htf}
Erd\'elyi A (ed) 1953
{\it Higher Transcendental Functions (The Bateman Manuscript Project)}
(New York: McGraw-Hill)

\bibitem{selberg} Forrester P J and Warnaar S O 2008
The importance of the Selberg integral
{\it Bull. Amer. Math. Soc.} {\bf 45} 489

\bibitem{PRAMajumdar} Dean D S, Le Doussal P, Majumdar S N and Schehr G 2016
Non-interacting fermions at finite temperature in a d-dimensional trap: Universal correlations
{\it Phys. Rev. A} {\bf 94} 063622

\bibitem{andrei} Andr\'eief C 1883
Note sur une relation entre les int\'egrales d\'efinies des produits des fonctions
{\it M\'em. Soc. Sci. Nat. Math. Bordeaux} {\bf 2} 1

\bibitem{wbarnes} Wikipedia, The Free Encyclopedia
https://en.wikipedia.org/wiki/Barnes\_G-function

\bibitem{nist} NIST {\it Digital Library of Mathematical Functions} http://dlmf.nist.gov

\bibitem{oeis} OEIS {\it The On-Line Encyclopedia of Integer Sequences} http://oeis.org

\end{thebibliography}
\end{document}